\documentclass[aps,prd,twocolumn,preprintnumbers,superscriptaddress,amssymb,eqsecnum,showpacs,showkeyes,secnumarabic,graphics,floatfix,nofootinbib,tightenlines,longbibliography]{revtex4-1}
\usepackage{graphicx}
\usepackage{bm}
\usepackage{cancel}
\usepackage{epsfig}
\usepackage{dcolumn}
\usepackage{amsmath}
\usepackage{hhline}
\usepackage{enumerate}
\usepackage[utf8]{inputenc}
\usepackage{ulem}
\usepackage[dvipsnames]{xcolor}
\usepackage[breaklinks=true,colorlinks=true,
linkcolor=blue,urlcolor=Blue,citecolor=MidnightBlue,
bookmarks=true,bookmarksopenlevel=2]{hyperref}
\usepackage{bm}
\usepackage{amsmath}
\usepackage{amssymb}
\usepackage{longtable}

\def\apjs{{Astroph.\@ J.\@ Suppl.\ }}

\def\mnras{{Mon.\@ Not.\@ Roy.\@ Ast.\@ Soc.\ }}

\def \beq  {\begin{equation}}
\def \eeq  {\end{equation}}
\def \ber  {\begin{eqnarray}}
\def \eer  {\end{eqnarray}}

\def \om    {\Omega}

\def \om0m {\Omega_{0\rm m}}

\usepackage{natbib}
\usepackage{graphicx}
\begin{document}

\newcommand{\newc}{\newcommand}
\newc{\be}{\begin{equation}}
\newc{\ee}{\end{equation}}
\newc{\D}{\partial}
\newc{\ie}{{\it i.e.} }
\newc{\eg}{{\it e.g.} }
\newc{\etc}{{\it etc.} }
\newc{\etal}{{\it et al.}}
\newc{\lcdm}{$\Lambda$CDM }
\newc{\lcdmnospace}{$\Lambda$CDM}
\newc{\wcdm}{wCDM }
\newc{\omom}{$\Omega_{0m}$ }
\newc{\omomnospace}{$\Omega_{0m}$}
\newc{\plcdm}{Planck18/$\Lambda$CDM }
\newc{\plcdmnospace}{Planck18/$\Lambda$CDM}
\newc{\wlcdm}{WMAP7/$\Lambda$CDM }
\newc{\fs}{{\rm{\it f\sigma}}_8}
\newc{\fsz}{{\rm{\it f\sigma}}_8(z)}
\newc{\geffz}{$G_{\rm eff}(z)$ }
\newc{\geffznospace}{$G_{\rm eff}(z)$}
\newc{\geff}{$G_{\rm eff}$ }

\newcommand{\nn}{\nonumber}
\newc{\ra}{\Rightarrow}
\title{Constraining a late time transition of $G_{\rm eff}$ using low-z galaxy survey data}
\author{G. Alestas}\email{g.alestas@uoi.gr}
\author{L. Perivolaropoulos}\email{leandros@uoi.gr} 
\affiliation{Department of Physics, University of Ioannina, 45110 Ioannina, Greece}
\author{K. Tanidis}\email{tanidis@fzu.cz}
\affiliation{CEICO, FZU---Institute of Physics of the Czech Academy of Sciences,\\Na Slovance 1999/2, 182 21 Prague, Czech Republic}

\date{\today}

\begin{abstract}
It has recently been pointed out that a gravitational transition taking place at a recent redshift $z_t$, reducing the effective gravitational constant $G_{\rm eff}$ by about $10\%$ for $z>z_t$, has the potential to lead to a resolution of the Hubble tension if $z_t\lesssim 0.01$.
Since $H(z)^2\sim G_{\rm eff}$, such a transition would also lead to sharp change of the slope of the Hubble diagram at $z=z_t$ and a sharp decrease in the number of galaxies per redshift bin at $z_t$. Here we attempt to impose constraints on such a transition by using two robust  low-z redshift survey datasets ($z<0.01$), taken from the Six-degree Field Galaxy Survey (6dFGS) as well as the 2MASS Redshift Survey (2MRS). In both surveys, we bin the data in redshift bins and focus on the number of galaxies in each bin ($\Delta N(z_i)$). We observe a peak in the distribution of galaxies near a distance of approximately 20 Mpc in both datasets. This feature could be attributed to galactic density fluctuations, to coherent peculiar velocities of galaxies  or to an ultra late-time gravitational transition in the same era. In the context of the later scenario we show that this feature could have been induced by a sharp change of $G_{\rm eff}$ by $\Delta G_{\rm eff}/G_{\rm eff} \simeq 0.6$ at $z_t\simeq 0.005$. Thus, in a conservative approach, this method can be used to impose constraints on a possible abrupt change of the gravitational constant taking place at very low redshifts.

\end{abstract}
\maketitle
\section{Introduction}
In recent studies \cite{Marra:2021fvf, Alestas:2020zol, Alestas:2021luu, Alestas:2021xes}, a gravitational transition has been proposed as a resolution of the Hubble tension \cite{Riess:2021jrx, Planck:2018vyg, Bernal:2016gxb, Verde:2019ivm, DiValentino:2021izs, Perivolaropoulos:2021jda, CANTATA:2021ktz, Vagnozzi:2021tjv, Dainotti:2021pqg}. In the context of this approach to the Hubble tension an abrupt $10\%$ change of the gravitational constant $G_{\rm eff}$ takes place at a redshift $z_t\lesssim 0.01$ such that gravity is about $10\%$ weaker at early times. This shift would change the Chandrasekhar mass $m_{Ch}\sim G_{\rm eff}^{-3/2}$ and thus the intrinsic luminosity of SnIa \cite{Amendola:1999vu, Gaztanaga:2001fh}, thus making them dimmer after the transition and changing their absolute magnitude $M$ by $\Delta M \simeq 0.2$ (the magnitude shift required for the resolution of the Hubble tension \cite{Alestas:2020zol, Camarena:2021jlr}).

Such a profound event could have signatures in a wide range of astrophysical data including effects on the history of the solar system, Cepheid standard candle data, Tully-Fisher data and galaxy redshift survey data. At the level of the solar system the well known increase of the number of impactors on the Moon and Earth surfaces during the last $100\; \rm Myrs$ ($z<0.007$) \cite{Bottke2007Sep} may be consistent with this type of gravitational transition. In addition interesting related features consistent with such a transition have been found recently in Cepheid \cite{Perivolaropoulos:2021jda, Mortsell:2021nzg} and Tully-Fisher data \cite{Alestas:2021nmi}. 

An abrupt shift of $G_{\rm eff}$ would also result in a corresponding shift of the Hubble expansion rate. Such a shift would be hard to detect directly at such low redshifts as $z<0.01$ but could be detectable indirectly as a signal in the observed number of galaxies per redshift bin at $z<0.01$. It would therefore be interesting to quantify the expected form of this signal and search for it in existing redshift survey data. This is the goal of the present analysis.

In addition to the effects of a gravitational transition on the background expansion rate investigated in the present analysis, the dynamical evolution of cosmological systems would also be affected. The corresponding effects of a $G_{\rm eff}$ transition by up to 10\% on the particle dynamics, is also an interesting issue that deserves investigation. However, in the present analysis we focus on the effects of a gravitational transition on the background evolution (Hubble expansion rate) and not on its effects on the particle dynamics.  The investigation of the effects of a transition on cosmological system dynamics at various scales is a separate and very interesting issue which is beyond the scope of the current analysis. For example the stability of N body systems under an abrupt change of the strength of gravity is highly non-trivial because N-body systems are chaotic and their stability is not easy to investigate with numerical methods. Even systems as small as the solar system have a Lyapunov time of more than $10\; \rm Myrs$ \cite{Murray:1999zd} which implies that the time-scale of any such instability would be similar to the time in the past when a such a possible destabilizing effect would have taken place. Thus for such late time events it is unlikely that an instability would be manifest by the present time.

The structure of this paper is the following: In the next section we specify the form of the expected signal in the detected number of galaxies $\Delta N(z)$ in each redshift bin of width $\Delta z$ centered around $z$. In section 3 we search for such a signal in the data of two galaxy surveys: The 6dFGS and the 2MRS. We then compare the 2MRS dataset with corresponding simulated catalogs based on the standard \lcdmnospace. Finally in section 4 we conclude, summarize our basic results and discuss possible future extensions of this analysis.

\section{Effects of a gravitational transition on redshift survey data}
In the context of the scalar-tensor modified gravity theories the gravitational constant acquires dynamical properties and thus the Friedman equation in redshift space may be expressed as 
\be
H(z)^2=\frac{8\pi G_{\rm eff}(z)}{3} \rho_{tot}
\label{freq}
\ee
where $\rho_{tot}$ refers to the total energy density including matter and an effective geometric dark energy component induced eg by the non-minimally coupled scalar field. Also, $G_{\rm eff}$ is the dynamical gravitational constant which is proportional to the inverse non-minimal coupling function $F(\Phi(z))$ of the scalar-tensor theory. The dynamical evolution of $G_{\rm eff}$ is severely constrained by a wide range of experiments and astronomical observations which constrain the time and redshift derivative of $G_{\rm eff}$ to $\frac{{\dot G}_{eff}}{G_{\rm eff}}<10^{-12}$ at various specific time ranges \cite{Uzan:2002vq,Will:2005va,Pitjeva:2021hnc} including the present time constrained mainly using solar system tests. Abrupt transitions of $G_{\rm eff}$ however can not be constrained by local constraints of the time derivative of $G_{\rm eff}$ since by definition, in the context of an abrupt transition $G_{\rm eff}$ would remain constant at (almost) all times. The overall change of $G_{\rm eff}$ between the present time and nucleosynthesis is weakly constrained to be less than about $10\%$ \cite{Alvey:2019ctk}.

Based on the generalized Friedman eq. (\ref{freq}) an abrupt change of $G_{\rm eff}$ at $z=z_t$ would also lead to a corresponding abrupt change of $H(z)$ such that
\be
\frac{\Delta G_{\rm eff}}{G_{\rm eff}} = 2 \frac{\Delta H}{H}
\label{deltagh}
\ee
In the Hubble flow $z_t>0.01$ such a transition is well constrained  by detailed Hubble diagram data based on Type Ia Supernovae (SnIa) \cite{Gaztanaga:2001fh}. For $z_t<0.01$ the Hubble diagram involves significant contributions from galactic density inhomogeneities and peculiar velocity effects and thus similar constraints are expected to be significantly weaker.

Using galaxy redshift surveys at $z<0.01$ it is possible to bin the observed galaxies in redshift bins of width $\Delta z$ such that there are $\Delta N(z_i)$ galaxies in the $i$ bin. In the presence of random peculiar velocities the measured redshift of a given galaxy may be written as
\be
cz = H_0 s + c\Delta z_r
\label{eq:cz}
\ee
where $H_0$ is the Hubble expansion rate at the galactic distance $s$ and $c\Delta z_r$ is a perturbation due to peculiar velocity effects and may be approximated to have random Gaussian distribution $(\mu=0,\,\sigma=300\, \text{km}\,\text{s}^{-1})$.

The number of galaxies that exist in a spherical shell with radius $s$ is given by,
\be
N(s)=\frac{4\pi}{3}s^3\rho(z)
\label{eq:Ns}
\ee
where we approximate the density at the redshift $\rho(z)=\rho_0(1+z)^3\approx\rho_0$ as homogeneous. The number of galaxies in the $i$ redshift bin may be easily obtained from eq. (\ref{eq:Ns}) as
\be
\Delta N(z_i)=4\pi \rho_0 \left(\frac{c}{H_0}\right)^3 (z_i-\Delta z_r)^2 \Delta z_i
\label{deltan}
\ee
where $\Delta z_i$ is width of the $i$ redshift bin assumed to be the same for all bins. Thus the predicted number of galaxies in the $i$ bin $\Delta N(z_i)$ is related to the number of galaxies in the $j=1$ bin as
\be
\Delta N(z_i) = \Delta N(z_1)\left(\frac{cz_i - c\Delta z_r}{cz_1 - c\Delta z_r}\right)^2 \left(\frac{H_{01}}{H_{0i}}\right)^3
\label{eq:DeltaN}
\ee
Violation of eq. (\ref{eq:DeltaN}) may be induced by either large density fluctuations of galaxies or coherent velocity flows. Eq. (\ref{eq:DeltaN}) however allows for a transition in the Hubble diagram slope $H_0$ at some redshift $z_t$. Such a transition could be expressed as
\be 
H_{0i} = H_{01} - \Delta H_0 \Theta(z_i-z_t)
\ee
In this case eq. (\ref{eq:DeltaN}) takes the form
\be
\Delta N(A,\delta,z_t,z_{i}) = A \left(\frac{cz_{i} - c\Delta z_{r}}{cz_{1} - c\Delta z_{r}}\right)^{2}[1 - \delta\; \Theta({z_{i}-z_t})]^{-3}
\label{eq:Nfun}
\ee
where $A\simeq \Delta N(z_{1})$ and $\delta \equiv \frac{\Delta H_0}{H_0}$ are parameters to be fitted by survey data.

\section{Analysis of Redshift Survey Data}
It is straightforward to implement the maximum likelihood method by minimizing $\chi^2$ with respect to the parameters $A$, $\delta\equiv \frac{\Delta H_0}{H_0}$ and $z_t$. Thus we minimize 
\be
\chi^{2}(A,\delta,z_t) = \sum^{N_{tot}}_{i=1} \frac{[\Delta N(z_{i})_{dat}-\Delta N(A,\delta,z_t,z_{i})]^2}{\sigma_{i}^{2}+\sigma_{s}^{2}}
\label{eq:chi2fun}
\ee
where $N_{tot}$ is the total number of bins,  $\sigma_{i}^{2} = N_{tot}/\Delta N(z_{i})_{dat}$ is the Poisson distribution error for each bin and $\sigma_{s}$ is the scatter error fixed such that the minimum $\chi_{min}^2$ is equal to one. Also $\Delta N(z_{i})_{dat}$ is the number of galaxies in each redshift bin after a  random perturbation $\Delta z_{r}$ is imposed on each measured galaxy redshift $cz_{i}$ to account for the random component $\Delta z_r$ in the parametrization (\ref{eq:Nfun}).

\subsection{The 6dF Galaxy Survey (6dFGS) $z<0.01$ subset}
\label{subsec:6dFGS}
\begin{figure*}
\centering
\includegraphics[width = 1 \textwidth]{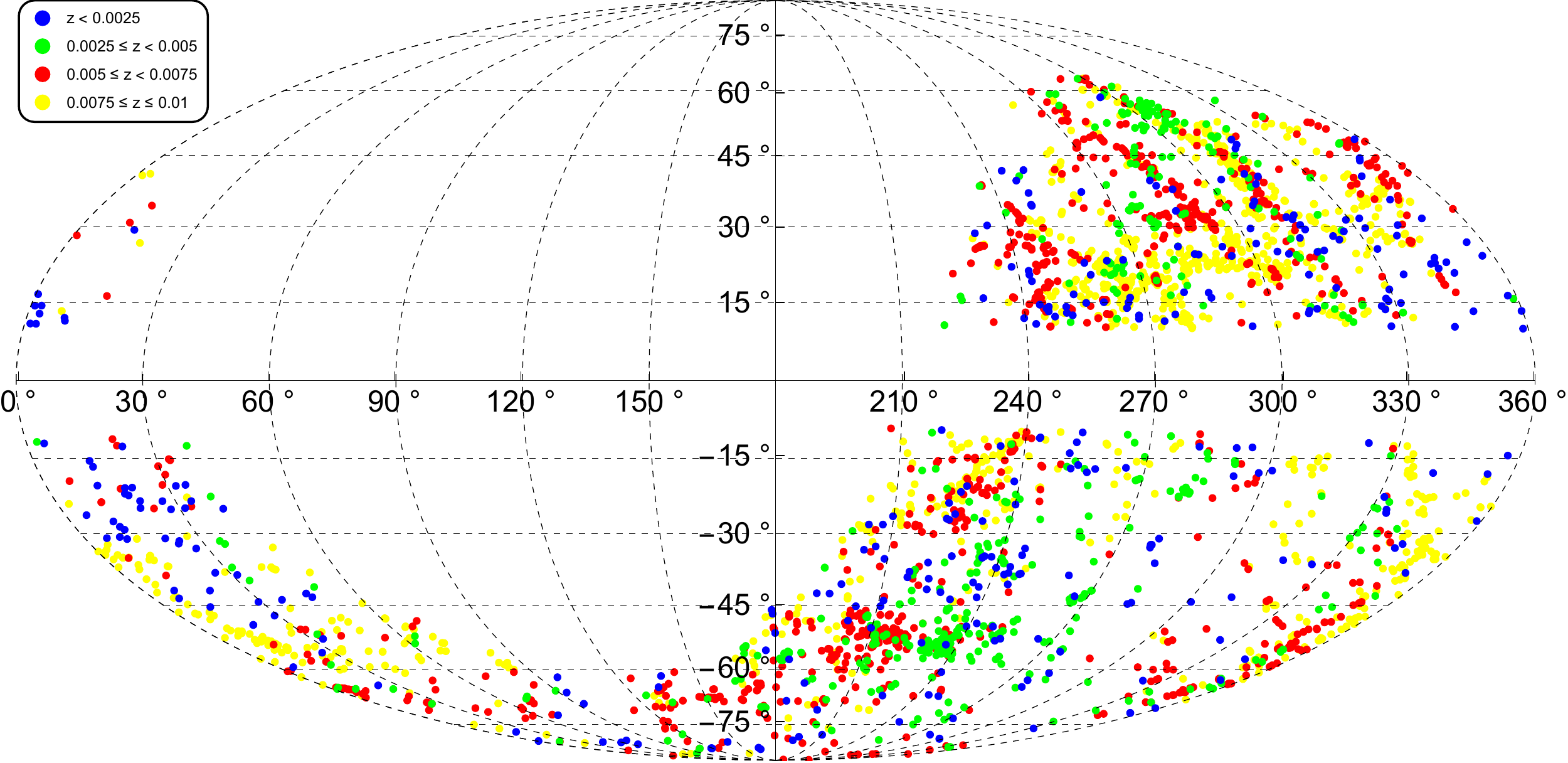}
\caption{The sky coverage of our collection of low-z 6dF data $(z<0.01)$ plotted in a Mollweide projection, using galactic coordinates. The data is split in four bins, of $\Delta z=0.0025$ increments. It is evident that data homogeneity is present when considering the different redshift increments.}
\label{skyplot}
\end{figure*}

\begin{figure*}
\centering
\includegraphics[width = 0.7 \textwidth]{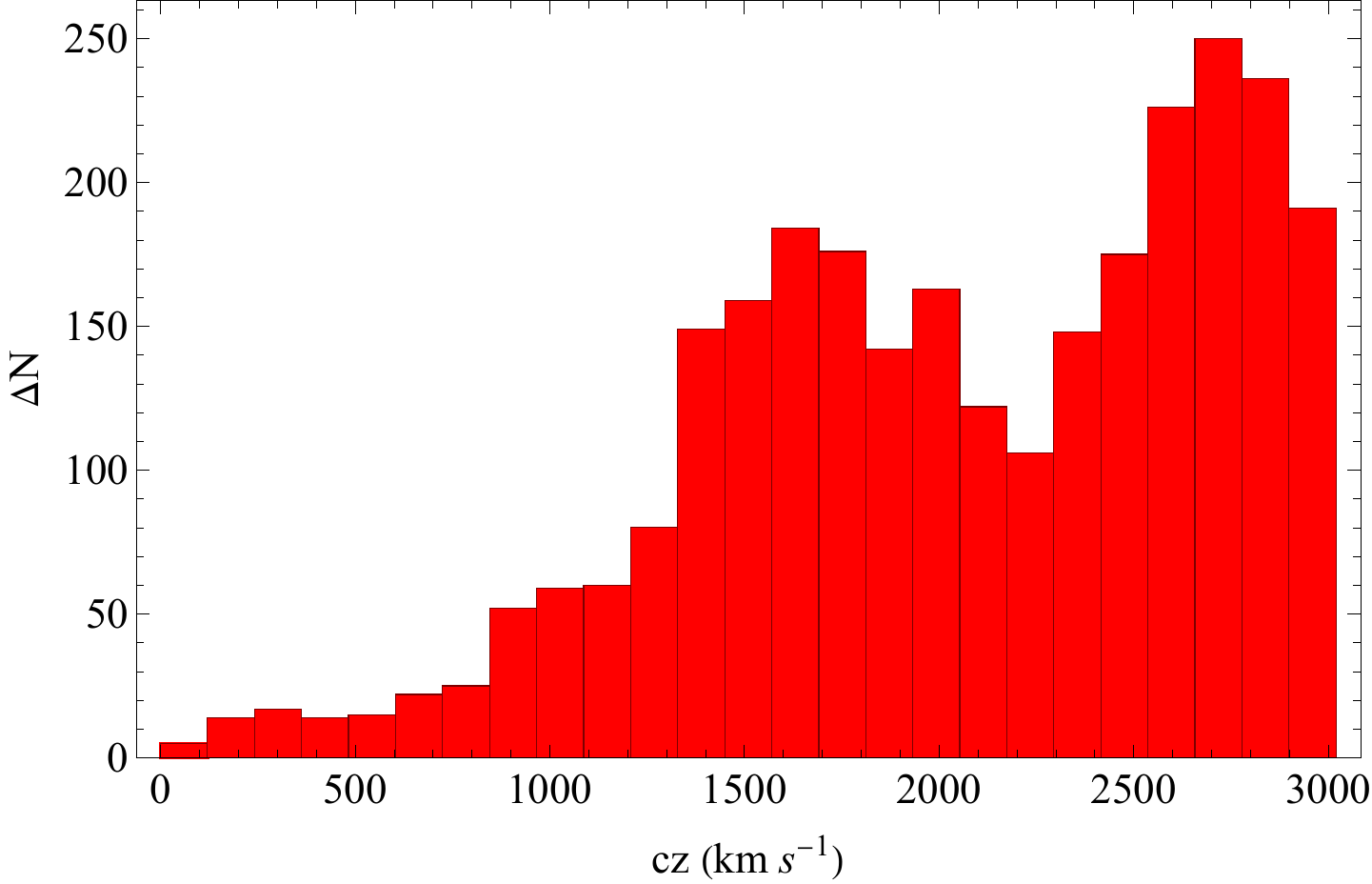}
\caption{The $\Delta N - cz$ histogram plot corresponding to the data of the 6dFGS dataset. It is evident that a large peak in the distribution of galaxies exists at $cz= 1500 -2000\, \text{km}\,\text{s}^{-1}$.}
\label{histogramplot}
\end{figure*}

We use the \href{http://www-wfau.roe.ac.uk/6dFGS/index.html}{6dFGS} \cite{2009MNRAS.399..683J, Jones:2005ya, Jones:2004zy, Beutler:2011hx, Beutler:2012px, Springob:2014qja, Johnson:2014kaa, Scrimgeour:2015khj}  focusing on the galaxies with $z<0.01$ ($\approx 2800$ galaxies). The peculiar velocity sample consists of 8885 galaxies in the Southern Hemisphere with $z<0.055$. The sky distribution of our low $z$ galaxy subsample is shown in Fig. \ref{skyplot} split in 4 redshift bins of $\Delta z=0.0025$ increments. The corresponding distribution of the galaxy sample in redshift space is shown in Fig. \ref{histogramplot} where  we split the sample in 25 redshift bins.

As shown in Fig. \ref{histogramplot}, there is a peak/dip feature in the redshift space number density of galaxies around $1500 -2000\, \text{km}\,\text{s}^{-1}$  ($21 - 28$ Mpc for a conservative value of $H_0 = 70\, \text{km}\,\text{s}^{-1}\,\text{Mpc}^{-1}$). This abrupt break in the redshift density is most probably due to density fluctuations of galaxies and/or coherent peculiar velocity flows. However, it may also be induced by a step-like transition of the Newton's constant $G_{\rm eff}$ occurring for $cz_t$ in the above range. Such gravitational transition would induce a similar transition in $H(z)\simeq H_0$ in accordance with eqs (\ref{freq}) and (\ref{deltagh}).  This type of transition could be expressed as \cite{Alestas:2020zol},
\be 
\mu_G(z)\equiv \frac{G_{\rm eff}}{G_{\rm N}} =[1+\Delta \mu_G \; \Theta (z-z_t)]
\label{eq:mutrans}
\ee
where $G_{\rm N}$ is the locally measured Newton's constant, and $\Theta(z)$ is the Heaviside step-function. This ansatz has been thoroughly explored in Refs. \cite{Alestas:2021luu, Marra:2021fvf}, where it is proposed that it would have the dual effect of solving both the Hubble and growth tensions. There have also been observational hints for such a transition at $\approx 20$ Mpc in Cepheid \cite{Perivolaropoulos:2021bds, Mortsell:2021nzg} and Tully-Fisher data \cite{Alestas:2021nmi}.

Assuming that the gravitational transition is the only cause of the observed dip in the $\Delta N(z)$ distribution we may use eq. (\ref{eq:Nfun}) to minimize $\chi^2$ (eq. (\ref{eq:chi2fun})) and thus obtain the best fit parameters $A$, $\delta$ and $z_t$. Such a fit for $\delta\equiv \Delta H_0/H_0$ should be interpreted as an upper bound for the transition amplitude $\delta$ and therefore also for the gravitational transition amplitude as obtained from eq. (\ref{deltagh}).

\begin{figure*}
\centering
\includegraphics[width = 1\textwidth]{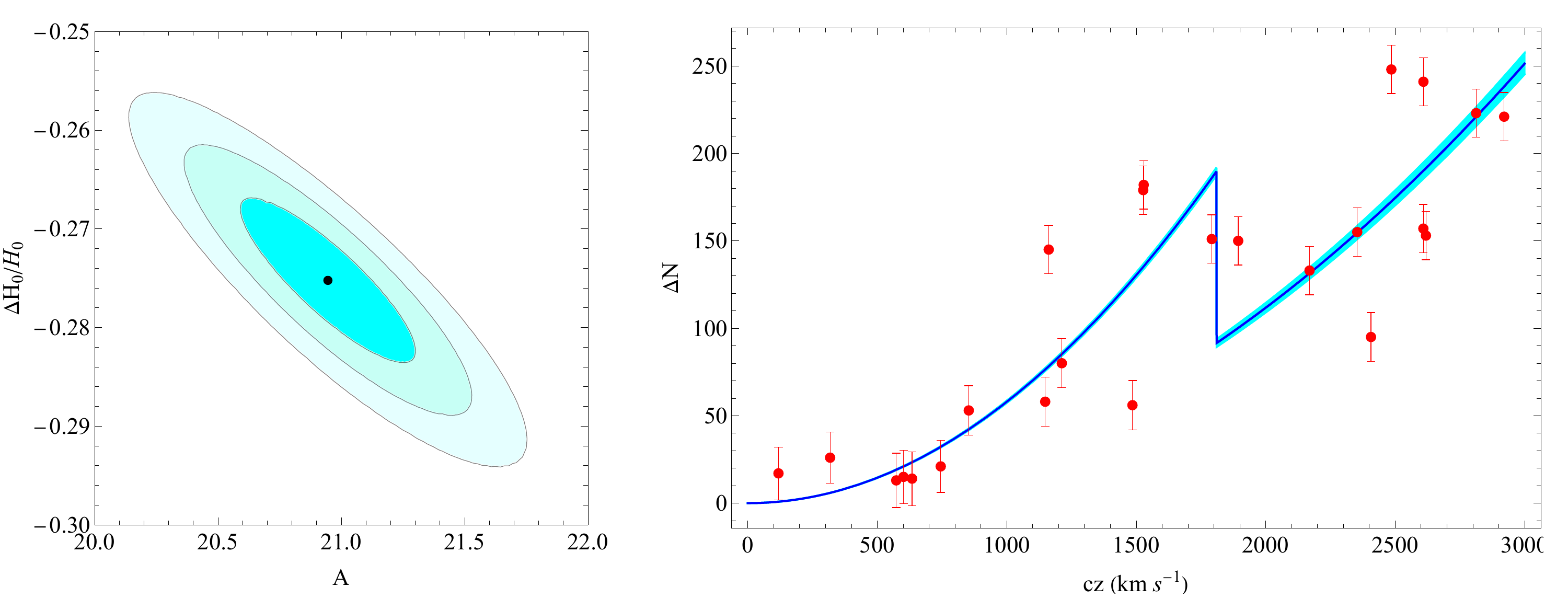}
\caption{{\it Left panel:} The $\Delta H_0 / H_0$ - $A$ likelihood contours corresponding from inner to outer to the 68\%, 95\% and 99.7\% C.L. intervals. The best-fit value is at the center (black bullet). The projected contours are taken at the best-fit value of the transition redshift $z_t$. {\it Right panel:} The $\Delta N$ - $cz$ plot of the best-fit (blue curve) of eq. \eqref{eq:Nfun} corresponding to the best-fit values of the parameters $\delta \equiv \Delta H_0 / H_0$, $A$ and $z_t$ along with the 68\% C.L. (cyan area). The binned numbers of galaxies in each redshift bin and their Poisson error are also shown (red errorbars).}
\label{dHdNdNczplot}
\end{figure*}

We thus obtain the best fit parameter values as $cz_t \approx 1810\pm 150\, \text{km}\,\text{s}^{-1}$, $A = 20.9 \pm 0.5$ and $\delta= \frac{\Delta H_0}{H_0} = -0.275\pm 0.01$ for a fixed value of $\sigma_s \approx 3.7$. In the left panel of Fig. \ref{dHdNdNczplot} we show the likelihood contours corresponding from inner to outer to the 68\%, 95\% and 99.7\% confidence level (C.L.) intervals in the parameter space $A - \delta$, while in the right panel we show the best fit (blue curve) form of  eq. \eqref{eq:Nfun} (with its 68\% C.L. error-band shown as cyan area) superposed with the $\Delta N(z_i)$ datapoints (red errorbars). In the context of the fit we have included the random gaussian perturbations of redshifts with $(\mu=0,\,\sigma=300\, \text{km}\,\text{s}^{-1})$ due to the effects of peculiar velocities in the data and have set $\Delta z_r=0$ in the ansatz (\ref{eq:DeltaN}), (\ref{eq:Nfun}). \footnote{This approach is equivalent to keeping the galaxy redshifts in their original form while including the random redshift perturbation in the ansatz (\ref{eq:DeltaN}). } 

We may therefore conclude that a possible transition of the Hubble diagram slope $H_0$ has to be smaller than the best fit value $\delta= \frac{\Delta H_0}{H_0} \leq -0.275\pm 0.01$. This upper bound for a Hubble diagram slope transition may be translated to an upper bound for an underlying gravitational transition using eq. (\ref{deltagh}) which leads to $\frac{\Delta G_{\rm eff}}{G_{\rm eff}}\lesssim 0.6$. Such an upper bound can easily accommodate the gravitational transition amplitude $\frac{\Delta G_{\rm eff}}{G_{\rm eff}}\simeq 0.1$ which has been proposed for the resolution of the Hubble and growth tensions \cite{Marra:2021fvf} which implies that this scenario remains viable in the context of the 6dFGS data.

\subsection{The 2MASS Redshift Survey (2MRS)  $z<0.01$ subset}
\label{subsec:2MRS}
\begin{figure*}
\centering
\includegraphics[width = 1 \textwidth]{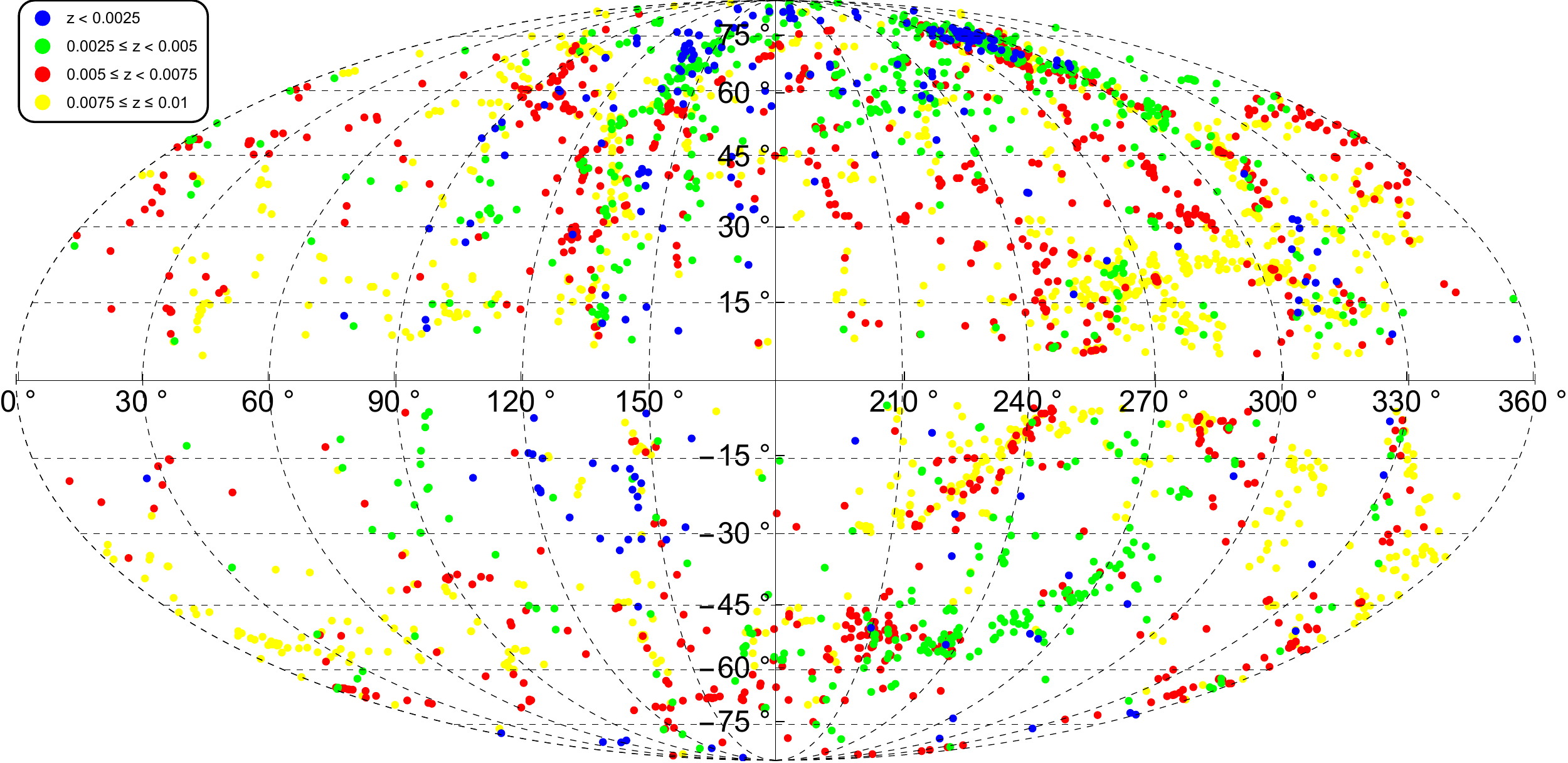}
\caption{The sky coverage of our collection of low-z 2MRS data $(z<0.01)$ plotted in a Mollweide projection, using galactic coordinates. The data is split into four bins, of $\Delta z=0.0025$ increments.}
\label{skyplot_2MRS}
\end{figure*}

We have repeated the analysis, in the same manner, using this time the larger and more recent \href{http://tdc-www.harvard.edu/2mrs/}{2MRS} \cite{2012ApJS..199...26H, Bilicki:2013sza, Alonso:2014xca, PierreAuger:2014yba, Tully:2015opa} dataset. This survey provides an almost full coverage of the sky ($\sim$70\%) including more data points than the 6dFGS peculiar velocity sample with a total number of 44599 spectroscopically observed sources at $z<0.15$ (the subsample with $z<0.01$ consists of $\approx 3200$ galaxies). We simulate the peculiar velocities in the data as described in \ref{subsec:6dFGS}. The sky distribution of these galaxies are shown in Fig. \ref{skyplot_2MRS} in four redshift bins up to $z=0.01$.

\begin{figure*}
\centering
\includegraphics[width = 0.7 \textwidth]{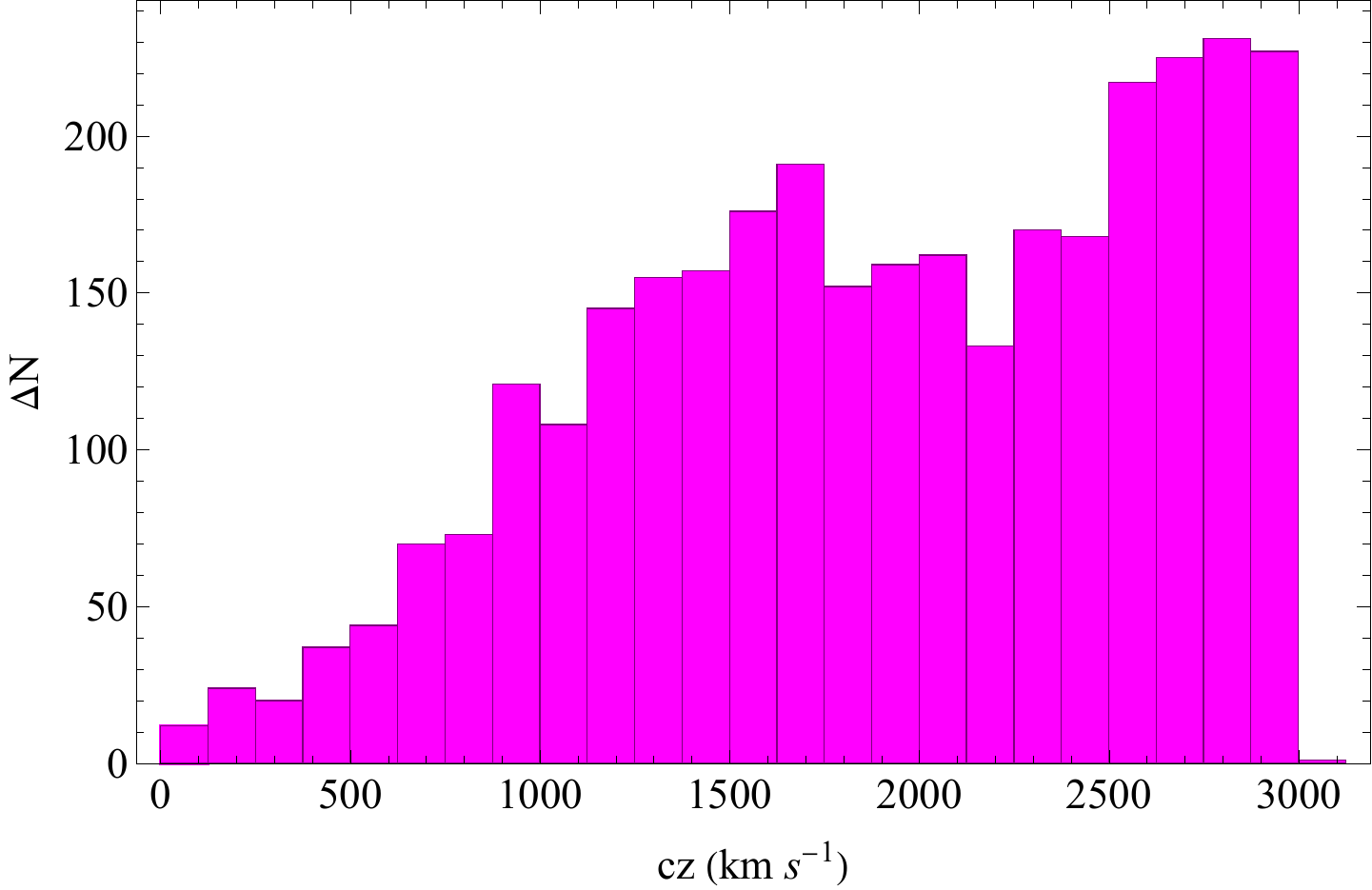}
\caption{The $\Delta N - cz$ histogram plot corresponding to the data of the 2MRS dataset. The same peak/dip feature in the distribution of galaxies as in the 6dFGS dataset exists at $cz= 1500 -2000\, \text{km}\,\text{s}^{-1}$, albeit it is less prominent.}
\label{histogramplot_2MRS}
\end{figure*}

The galaxy distribution in redshift space is shown in Fig. \ref{histogramplot_2MRS} where a similar dip may be seen at $cz \approx 1500\, \text{km}\,\text{s}^{-1}$.  As in the case of the 6dFGS dataset, this feature is most likely due to density variations of the galaxy distribution and to peculiar velocity flows. If however we assume that it is due to a gravitational transition of the form (\ref{eq:mutrans}) leading to a transition of the Hubble parameter, then we can derive an upper bound on $\frac{\Delta G_{\rm eff}}{G_{\rm eff}}$

\begin{figure*}
\centering
\includegraphics[width = 1\textwidth]{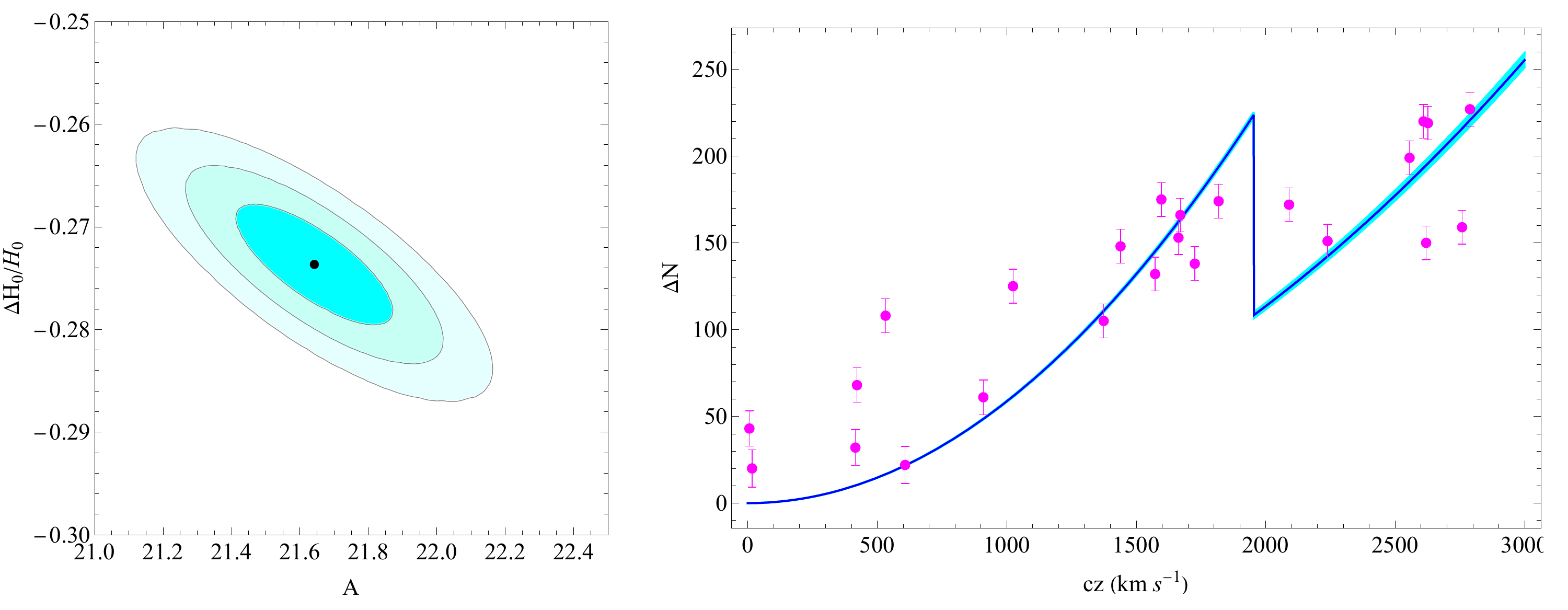}
\caption{{\it Left panel:} Same as Left panel of FIG. \ref{dHdNdNczplot} but for the 2MRS sample. {\it Right panel:} Same as Right panel of FIG. \ref{dHdNdNczplot} but for the 2MRS sample.}
\label{dHdNdNczplot_2MRS}
\end{figure*}

In this case, the best fit parameter values are similar as in the 6dFGS and take the form $A = 17.5\pm 0.5$, $\delta=\frac{\Delta H_0}{H_0} = -0.28\pm 0.01$ and $c z_t \approx 1783 \pm 150\; \text{km}\,\text{s}^{-1}$ for $\sigma_s \approx 3.4$. We have plotted the confidence contours in the  $A-\delta$ parameter subspace (left panel) as well as the best-fit form of $\Delta N(z)$ based on  eq. \eqref{eq:Nfun} (right panel),  in Fig. \ref{dHdNdNczplot_2MRS}.

\subsection{Comparison with Simulated Galaxy Surveys}

\begin{figure*}
\centering
\includegraphics[width = 0.7 \textwidth]{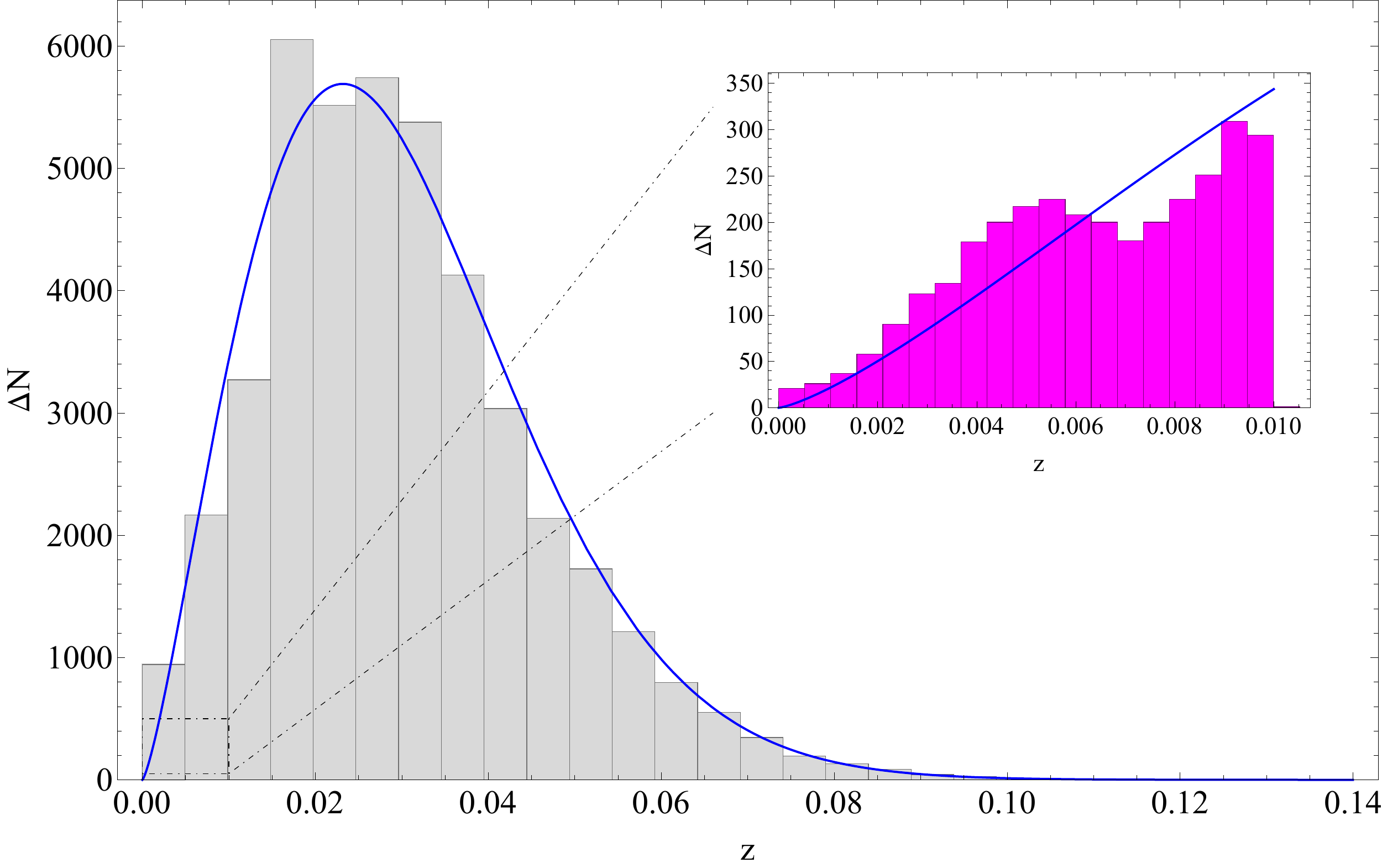}
\caption{The number of galaxies in each redshift bin for the entire 2MRS sample (lightgray histogram), as well as for our subsample with $z<0.01$ (magenta histogram), superimposed with the fitting function eq. \eqref{eq:2MRS}.}
\label{Distributionplot_2MRS}
\end{figure*}

In order to see how commonly, if at all, would the abrupt peak/dip in the galaxy distribution appear also in the simulated data based on a standard \lcdm cosmology, we have used the Cosmological Lofty Realizations (\href{https://github.com/damonge/CoLoRe}{CoLoRe}) \cite{Ramirez-Perez:2021cpq} software package that supports lognormal fields to generate synthetic realizations for the 2MRS galaxy survey. We opted to generate mock catalogs for this survey since as we saw in \ref{subsec:2MRS}, it is more complete than the 6dFGS peculiar velocity sample. In particular, for the simulated catalogs on top of the assumed standard \lcdm model for the input Gaussianised matter power spectrum $P(k)$ at $z=0$, we also included a constant galaxy bias with redshift $b(z)=1.3$ (a value found to be a good approximation at non-linear scales) and the approximated fitting function for the redshift distribution of the 2MRS found by \cite{Ando:2017} that reads:  

\begin{align}\label{eq:2MRS}
    \frac{dN}{dz}=\frac{N_g \beta}{z_0 \Gamma\left[(m+1)/\beta\right]}\left(\frac{z}{z_0}\right)^m \exp{\left[-{\left(\frac{z}{z_0}\right)}^\beta\right]}
\end{align}
with \(\beta=1.64\), \(z_0=0.0266\), \(m=1.31\) and the total number of sources \(N_g=44599\), see Fig. \ref{Distributionplot_2MRS}.

Then we set up 500 simulations with the aim to clarify if the peak/dip feature in $\Delta N(z)$ would occur naturally in them in the context of a standard \lcdm cosmology. In this case the best fit value of $\delta$ derived in the context of our analysis can only be viewed as an upper bound.

\begin{figure*}
\centering
\includegraphics[width = 1\textwidth]{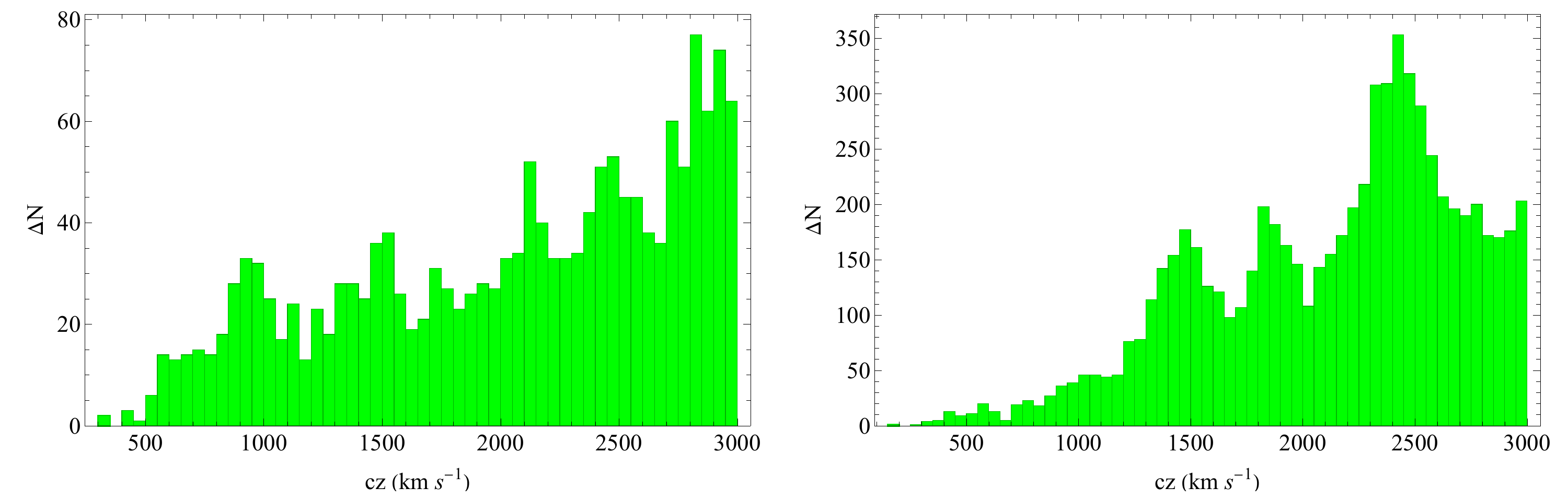}
\caption{The $\Delta N - cz$ histogram plot corresponding to the data of two random simulated datasets based on \lcdm. Features like the one shown in the real data appear to be common due to galactic number density inhomogeneities.}
\label{histogramplot_Sim}
\end{figure*}

As expected, the simulated data indicate that peaks and dips like those found in the real data occur commonly in the corresponding simulated datasets based on standard \lcdm cosmology due to density and peculiar velocity effects. This is demonstrated in the two simulated 2MRS datasets shown in  Fig. \ref{histogramplot_Sim} randomly chosen from the 500 mock catalogs to showcase here.  The magnitude of these features  overwhelms any possibility of interpreting the observed peak/dip feature in the real $\Delta N(z)$ data as a signature of the presence of a $\Delta H_0 / H_0$ transition and only allows the interpretation of the best fit value of $\Delta H_0 / H_0$ as a bound on the magnitude of a possible corresponding Hubble and gravitational transition.

\section{Conclusion}

We have analysed the low z distribution ($z<0.01$) of galaxies in the 6dFGS peculiar velocity sample and the complete 2MRS galaxy survey searching for a signal consistent with a gravitational abrupt transition. We have identified such a signal consistent with a gravitational transition $\frac{\Delta G_{\rm eff}}{G_{\rm eff}}\simeq 0.6$. Such a signal however is degenerate with corresponding expected features emerging due to density fluctuations in the number density of galaxies and peculiar velocity flows. This was demonstrated by simulating the expected redshift distribution of the galaxies for the more complete 2MRS catalog in the context of standard \lcdm without gravitational transition. Thus the detected signal can only be interpreted as an upper bound on the magnitude of such a gravitational transition. Even though this bound is weaker than corresponding bounds obtained using nucleosynthesis and CMB power spectrum data implying $\frac{\Delta G_{\rm eff}}{G_{\rm eff}}\lesssim 0.1$ it remains interesting for two reasons:
\begin{itemize}
    \item It is based on  a novel method for constraining a gravitational transition.
    \item It focuses on a very specific ultra-late redshift range
    \item It indicates that the proposed magnitude of a gravitational transition $\frac{\Delta G_{\rm eff}}{G_{\rm eff}}\simeq 0.1$ for the resolution of the Hubble and growth tension is consistent with current galaxy survey data.
\end{itemize}

An assumption used in our analysis is that of uncorrelated Gaussian random peculiar velocity field which was superposed in the Hubble velocity flow. This assumption ignores the local bulk flows and the possible correlation among the redshift bins in estimating the uncertainties. The use of a diagonal covariance matrix, instead of using the full covariance matrix is a simplification which we had to implement, since the detailed form of the velocity flows on the considered scales is not precisely known and thus we do not have access to a reliable and detailed form of the full covariance matrix. However, it is clear that the use of the full covariance matrix would weaken the constraint on $G_{\rm eff}$ derived here. We have verified this result using toy covariance matrices that fully correlates only neighboring redshift bins. Following this approach, we have observed a very small (few percent) increase in the uncertainties of the best-fit parameters. This indicates that even after the inclusion of velocity correlations and the full covariance matrix, the constraint would remain consistent with a $10\%$ gravitational transition at $z<0.01$. Thus our approach can indeed lead to new constraints on a gravitational transition at redshifts $z<0.01$ (last $150\; \rm Myrs$) but these constraints are not powerful enough to rule out the gravitational transition class of models for the resolution to the Hubble tension.

Interesting extensions of the present analysis include the following:
\begin{itemize}
    \item The identification of actual distances and peculiar velocities of the galaxies included in the considered surveys. In that way, it can be estimated to what extent the identified feature in $\Delta N(z)$ is due to galactic density variations and/or peculiar velocity flows. If this feature can not be fully explained as a density and peculiar velocity effect then it is possible that at least part of it may be due to a gravitational transition. 
    \item The identification of additional astrophysical datasets beyond galaxy surveys, Tully-Fisher data and solar system history which may lead to constraints on the magnitude of such a profound effect like an ultra-late gravitational transition.
    \item The simulation of the solar system evolution (and in particular of the Oort cloud) to identify the change of impactor rate in the context of a late gravitational transition. In this context, the solar system history could become a useful laboratory for the constraint of such a transition.
\end{itemize}

\textbf{Numerical Analysis Files}: The numerical files for the reproduction of the figures can be found \href{https://github.com/GeorgeAlestas/G_Constraints}{here}.

\section*{Acknowledgements} We thank  Prof. Avi Loeb for interesting discussions that lead to the idea of this project and Profs. David Alonso and Stefano Camera for their help in configuring scripts for the CoLoRe software. GA's research is supported by the project “Dioni: Computing Infrastructure for Big-Data Processing and Analysis” (MIS No. 5047222) co-funded by European Union (ERDF) and Greece through Operational Program “Competitiveness, Entrepreneurship and Innovation”, NSRF 2014-2020. KT is supported by the European Structural and Investment Fund and the Czech Ministry of Education, Youth and Sports (Project CoGraDS - CZ.02.1.01/0.0/0.0/15\_003/0000437).

\raggedleft
\bibliography{references}
\end{document}